\input AHTOH-E.STY
\let\+\oplus

\UDC{%
512.572
+519.516.5
+519.696
+512.554       
}

\grants{\RFBR 11-01-00945}

\title{
The identities of additive binary arithmetics
}
\author{
Anton A. Klyachko
\quad
Ekaterina V. Menshova
}
\address{\myAddress\quad ekaterina.menshova@gmail.com}

\abstract{%
\narrower
Operations of arbitrary arity expressible via addition
modulo $2^n$ and bitwise addition modulo $2$ admit a simple description.
The identities connecting these two additions have a finite basis.
Moreover, the universal algebra $\Z/2^n\Z$ with these two
operations is rationally
equivalent to a nilpotent ring and, therefore, generates a Specht
variety.
}

\s
0. Introduction

On the set of integers
$\{0,1,\dots,q-1\}=\Z_q$, where $q$ is
a power of two, we consider two natural operation: addition modulo~$q$
and bitwise addition modulo 2. In computer literature, these
operations are usually denoted by {\tt ADD} and {\tt XOR}; they are
hardware implemented in all modern computers, as far as we know.%
\fn{%
Usually, the command {\tt ADD}
is simply called addition, because it is used
mostly for obtaining the usual sum of positive integers; however,
actually the processor performs the addition modulo a large
number $q$
(for example, $q=2^{32}$ for 32-bit processors etc.).
}

We consider
two natural questions.
\item{1.}
{\sl What function $\Z_q^k\to\Z_q$ can be expressed via these two
operations?}
\item{2.}
{\sl What identities connect these two operations?}

\enditem
Theorem 1 gives a complete answer to
the first question; we obtain a simple fast algorithm
deciding
whether or not an arbitrary given function can be expressed
via {\tt ADD} and {\tt XOR}. We also calculate the total number
of $k$-argument functions expressible via these two operation
(Corollary 1).

We do not give an explicit answer to the second question, but we
prove that, for each $q$, all identities connecting {\tt ADD} and
{\tt XOR} follow from a finite number of such identities (Theorem 2)
and there exists an algorithm writing down such a finite basis of
identities for any given $q$ (Corollary 2).

The problem of existence of finite basis of identities was extensively
studied for groups, semigroups, rings, linear algebras (see, for example,
[BaOl88], [Neum69], [Belo99], [VaZe89], [Grish99], [Zaits78], [Keme87],
[Kras90], [Laty73], [Lvov73], [Olsh89], [Shch99], [GuKr03], [Kras09],
[Speht52]
and literature cited therein), but the ``applied" algebra with
operations~{\tt ADD}~and~{\tt XOR} has never been studied from this
point of view, as far as we known.

In algebraic term, Theorem 1 is an explicit description
of the
free algebras of the variety, generated by the algebra~$\Z_q$ with two
binary operations~{\tt ADD}~and~{\tt XOR}; Theorem 2 says that
this variety is finitely based (i.e. it has a finite basis of
identities).
See, for example, [BaOl88] or [GA91] for
necessary information about varieties of universal
algebras.

\noindent
Most of {\bf Notation}
we use is standard. Note only that the
addition modulo $q$ (i.e. {\tt ADD}) is denoted by
$+$; the bitwise addition modulo 2 (i.e.~{\tt XOR}) is
denote by $\+$. The symbol $a_i$
denotes the $i$th bit of a number $a\in\Z_q$; the bits with negative
numbers are assumed to be zero.
The set $\{0,1,\dots,q-1\}=\Z_q$
when considered as a universal algebra with operations $+$ and $\+$ is
denoted by the symbol $A_q$. The multiplication by integer numbers in
$A_q$ is always considered as multiplication modulo $q$. These
multiplications are obviously expressed in terms of addition~$+$,
for example,
$
3x=x+x+x \hbox{ and }
-3x=(-3)x=-(3x)=\underbrace{x+x+\dots+x}_{3(q-1)\; terms}.
$

The authors thank
an anonymous referee
for a lot of useful remarks and pointing out a flaw in the original
proof of the main theorem (the easier direction).
We also thank A.~E.~Pankratiev for useful comments.

\s
1. Definitions and results

A function $f\:A_q^k\to A_q$ is called \emph{algebraic} if
it can be
expressed via operations $+$ and $\+$. More precisely,
the set $F_{k,q}$
of algebraic  $k$-argument functions is the
inclusion-minimal set
of functions, satisfying the following
conditions:
\item{1)}
the functions $f(x,y,\dots)=x,\ f(x,y,\dots)=y,\dots$
belong to $F_{k,q}$;
\item{2)}
if functions $f$ and $g$ belong to $F_{k,q}$, then the functions
$f+g$ and $f\+g$ belong to $F_{k,q}$.

\enditem
The set $F_{k,q}$ of all
algebraic $k$ argument functions forms a
universal
algebra with respect
to
operations $+$ and $\+$; this is the \emph{free algebra of rank $k$
of the
variety generated by the algebra $A_q$}.

\Th 1.
A function $f\:A_q^k\to A_q$ is algebraic if and only if,
for any $i$,
the $i$th bit of its value is expressed via bits of
the arguments by a
formula of the form
$$
(f(x,y,\dots))_i=
g(x_i,y_i,\dots;x_{i-1},y_{i-1},\dots;x_{i-2},y_{i-2},\dots;\dots)
\eqno{(*)}
$$
\(the bits with negative numbers are assumed to be zero\),
where $g$ is an independent of $i$
Zhegalkin polynomial \(over~$\Z_2$\) without free term, whose weight
does not exceed 1.

The \emph{weight} or the \emph{reduced degree} of a polynomial
in variables
$x_i,y_i,\dots,x_{i-1},y_{i-1},\dots,x_{i-2},y_{i-2},\dots$ is the
maximal weight of its monomials; the weight of a
monomial is the sum of weights of its
variables; the weight of
variables~$x_{i-l},y_{i-l},\dots$ is the number $2^{-l}$.
(Here, $i$ is a formal parameter.)

\Example 1.
If $q=8$ and $k=1$, then there are exactly four
Zhegalkin monomials whose
weights does not exceed one:
$x_i$ (weight 1), $x_{i-1}$ (weight $1\over2$),
$x_{i-2}$ (weight $1\over4$), and
$x_{i-1}x_{i-2}$ (weight $3\over4$).
(Here, we use that no variable occurs in a Zhegalkin
monomial more than once.)
Therefore, there are $2^4$
polynomials of weight not exceeding one.
Thus, the algebra~$F_{1,8}$ consists of 16 elements.
For example, the algebraic function corresponding to the
Zhegalkin polynomial
$x_i\+x_{i-1}x_{i-2}$ has the form
$$
f(x)=
f(x_0,\;x_1,\;x_2)=
(x_0\+x_{-1}x_{-2},\;x_1\+x_{0}x_{-1},\;x_2\+x_{1}x_{0})=
(x_0,\;x_1,\;x_2\+x_{1}x_{0})
$$
(because the bits with negative numbers are zero).
In other words,
$$
f(0)=0,\quad
f(1)=1,\quad
f(2)=2,\quad
f(3)=7,\quad
f(4)=4,\quad
f(5)=5,\quad
f(6)=6,\quad
f(7)=3.
$$

Theorem 1 makes it possible to construct the following simple

\smallskip\noindent
{\bf ALGORITHM} determining whether or not a given function
$f\:\Z_{2^\kappa}^k\to\Z_{2^\kappa}$ is
algebraic (i.e. can be expressed via {\tt ADD}
and~{\tt XOR}).
\item{1.}
Write the most significant bit $(f(x,y,\dots))_{\kappa-1}$ of the value
of the
function $f$ as a Zhegalkin polynomial
\newline
$g_{\kappa-1}(x_0,y_0,\dots,x_1,y_1,\dots)$
in bits of arguments and verify that the
weight
of this polynomial (for $i=\kappa-1$)
does not
exceed one and the free term is zero.
If the weight is higher
or the free term is nonzero, then
exit the
program with the answer {\tt NO}.

\item{2.}
Make the following substitutions in the polynomial $g_{\kappa-1}$:
$$
x_0\to 0,\ x_1\to x_0,\ x_2\to x_1,
\dots,\ x_{\kappa-1}\to x_{\kappa-2},\
y_0\to 0,\ y_1\to y_0,\ y_2\to y_1,
\dots,\ y_{\kappa-1}\to y_{\kappa-2},\dots
\eqno{(**)}
$$
and verify that the obtained polynomial $g_{\kappa-2}$ coincides
with the
polynomial giving $(\kappa-2)$-th bit of the function~$f$.
If not, then exit the program with the answer {\tt NO}; if yes, then
continue.
$$
\dots\quad\dots\quad\dots\quad
$$
\item{$\kappa-1$.}
Make substitutions $(**)$ in the polynomial $g_2$
and verify that the obtained polynomial $g_1$ coincides
with the
polynomial giving the 1st bit of the function $f$.
If no, then exit the program with the answer {\tt NO}; if yes, then
continue.

\item{$\kappa$.}
Make substitution $(**)$ in the polynomial $g_{1}$
and verify that the obtained polynomial $g_{0}$ coincides with
the
polynomial giving the least significant bit of the function $f$.
If no, then exit the program with the answer {\tt NO}; if yes, then
exit the program with the answer {\tt YES}.

\medskip
\noindent
It is clear that this algorithm can be easily made uniform with
respect to $\kappa$.

For example, the function of multiplication of two numbers modulo $q$
cannot be
expressed via {\tt ADD} and {\tt XOR} (the algorithm stops at the
first step because of the condition on the weight); this is not
surprising, of course.  However, the one-argument function
$x\mapsto xy$
is algebraic for each given $y\in\Z_q$, as was already mentioned.

The proof of Theorem 1 is constructive and gives some
algorithm making it possible to express a given function~$f$ via
{\tt ADD} and {\tt XOR}
(if it is expressible), but this algorithm is much more
complicated.

Example 1 can be easily generalised. Calculating monomials for any
$q$ and $k$, we
obtain the following assertion.

\Corollary 1.
The free algebra $F_{k,q}$ comprises
$$
2^{{1\over k!}
\({q\over2}+1\)\({q\over2}+2\)\dots\({q\over2}+k\)
-1}
\eqno{(1)}
$$
elements.

\Proof
In the case where $k=1$, there are precisely ${q\over2}$ monomials of
weight at most one. Indeed, by virtue of the uniqueness of
binary decomposition of an integer, there exists precisely one monomial
of each weight $s\cdot{2\over q}$, where $s\in\{1,2,\dots,{q\over2}\}$.
Namely, this is the
monomial
$$
x_{i-l_1}x_{i-l_2}\dots x_{i-l_p},
\qqbox{where}
s=
2^{\kappa-1-l_1}+2^{\kappa-1-l_2}+\dots+2^{\kappa-1-l_p},
\qbox{and $2^\kappa=q$}.
$$
This implies that, for any integer $k$,
the number of monomials of weight at most one
coincides with the number of nonzero tuples of nonnegative integers
$(n_1,\dots,n_k)$ with sum at most $q\over2$
(here, $n_i\cdot{2\over q}$ is the weight with respect
to the $i$th variable).
It is well known that
the number
of such tuples is
$$
{({q\over2}+1)({q\over2}+2)\dots({q\over2}+k)
\over
k!}
-1.
$$
Therefore, the total number of polynomials of weight at most one
can be found by formula (1).

\smallskip

The following assertion is a reformulation of Theorem 1.

\Th 1$'$.
A function $f\:A_q^k\to A_q$ is algebraic if and only if it
can be written in the form
$$
f(x,y,\dots)=\bigoplus_{i}
\((2^{k_{i,1}}x)\odot(2^{k_{i,2}}x)\odot
\dots\odot
(2^{l_{i,1}}y)\odot(2^{l_{i,2}}y)\odot
\dots\),
$$
where the inequality
$
2^{-k_{i,1}}+2^{-k_{i,2}}+\dots+2^{-l_{i,1}}+2^{-l_{i,2}}+\dots\le1
$
holds for each $i$.

Henceforth, the symbol $\odot$ denotes the bitwise multiplication
modulo 2 (conjunction).

\medskip

As for identities, we note first that, with respect to
each of the operations $+$ and $\+$, the algebra $A_q$ is
an abelian group of exponent $q$ and 2, respectively.
Therefore, all identities involving only one of these two operations
follow from the identities
$$
(x+y)+z=x+(y+z),\ x+qy=x,\ x+y=y+x, \quad
(x\+y)\+z=x\+(y\+z),\ x\+(y\+y)=x,\ x\+y=y\+x.
$$
Identities involving the both operations are more complicated.
The simplest example of such
identity is $qx=x\+x$ which expresses the coincidence of the
zero elements of these two group structures.
A less trivial example is
${q\over2}(x+y)={q\over2}(x\+y)$ (this identity expresses the
coincidence of the
additions $+$ and $\+$ at the least significant bit).

\Th 2.
For any integer power of two $q$, the algebra $A_q$ has a finite
basis of
identities. Moreover, the algebra $A_q$ generates a Specht variety.%
\fn{%
This means that any algebra of signature $(+,\+)$
satisfying all identities of the algebra $A_q$ has a finite
basis of
identities.
}

The finiteness of an algebra per se does not implies the finiteness
of a basis of its identities. A finite basis
of identities exists for each finite
group [OaPo64] (see also [Neum69]), each finite
associative or Lie ring
([Lvov73], [Kruse73], [BaOl75]), but not for each finite semigroup and
not for
each finite ring (see~\hbox{[BaOl88]}).

To prove Theorem 2, we use
well-known
nilpotency arguments rather than the finiteness. It is known
that a finite basis of identities exists for any nilpotent ring (i.e. a
ring in which all sufficiently long products vanish) and any nilpotent
group (i.e.  a group in which all sufficiently long multiple commutators
equal to one) (see~[Neum69]). The algebra $A_q$ is neither a group nor a
ring.  However, it turns out that this algebra is \emph{rationally
equivalent} (in the sense of Mal'tsev) to a nilpotent ring, i.e. the
algebra~$A_q$ can be endowed with a structure of nilpotent ring in such a
way that the addition and multiplication of the ring can be expressed via
operations $+$ and $\+$ and vice versa: the operation $+$ and $\+$ can be
expressed via the addition and multiplication of the ring.

\Th 3.
The algebra $A_q$ is rationally equivalent to a nilpotent commutative
nonassociative ring $(\Z_q,\+,\o)$. The addition $\+$ is the usual
bitwise addition modulo two, the multiplication $\o$ is defined by the
formula $x\o y=2(x\odot y)$, where $\odot$ is the bitwise
multiplication modulo two (conjunction), and the multiplication by two is
the multiplication by two modulo $q$, i.e. the shift of digits.

In the following section, we prove Theorem 1. In section 3, we prove
Theorem 3, which immediately implies Theorem 2, because
any nilpotent ring has a finite basis of
identities (and generates a Specht variety).

\s
2. Proof of Theorem 1

The element $[x,y]\:=x \oplus y \oplus (x+y)$
is called
the \emph{commutator} of the elements $x,y\in A_q$.
The commutator is the difference between the sum $\+$ and the sum $+$
of two elements; the $i$th bit of the commutator $[x,y]$ is the
carry to the $i$th digit during
execution of
the standard addition algorithm for $x+y$.

The following lemma is well known and widely used in electronic adders.

\Proposition 1.
The bits of the commutator
satisfies the equality
$$
[x, y]_i = x_{i-1}y_{i-1} \oplus
[x, y]_{i-1}(x_{i-1}\oplus y_{i-1}).
\eqno{(2)}
$$

\Proof
The carry $c_i=[x,y]_i$ to the $i$th bit is formed as follows:
$$
c_i=\cases{
1,& if among three bits $x_{i-1}, y_{i-1}, c_{i-1}$
the majority (i.e. two or three) is ones;
\cr
0,& if among three bits $x_{i-1}, y_{i-1}, c_{i-1}$
the majority is zeros.
\cr
}
$$
The Zhegalkin polynomial for this Boolean function is
$$
c_i=x_{i-1}y_{i-1}\+y_{i-1}c_{i-1}\+c_{i-1}x_{i-1}=
x_{i-1}y_{i-1}\+c_{i-1}(x_{i-1}\+y_{i-1}),
$$
as required.

\smallskip

Formula (2) can be rewritten in the form
$
[x,y] = 2(x\odot y\+[x,y]\odot(x\+y))
$
or
(applying distributivity of the multiplication by two
with respect to $\odot$ and $\+$ and distributivity
of
$\odot$ with respect to $\+$) in the form
$$
(2x)\odot(2y)=[x,y]\+(2[x,y])\odot(2x)\+(2[x,y])\odot(2y).
\eqno{(2')}
$$
Using
formula (2), it is easy to show that the
$i$th bit of the sum $x+y=x\+y\+[x,y]$
can be evaluated as
$$
(x+y)_i=
x_i\oplus y_i\oplus x_{i-1}y_{i-1}\oplus x_{i-1}x_{i-2}y_{i-2}\oplus
y_{i-1}x_{i-2}y_{i-2}\oplus \dots=
the\ sum\ of\ all\ monomials\ of\ weight\ 1.
\eqno{(2'')}
$$

Now, we proceed to prove Theorem 1.
Let $M$ be the set of all functions $A_q^k\to A_q$ of the
form $(*)$.
We have to prove two assertions.
\item{1.}
Any function from $F_{k, q}$
belongs to $M$;
\item{2.}
Any function from $M$ belongs to
$F_{k, q}$.

\enditem
The first assertion is easy to verify.
The functions $f(x,y,\dots)=x,\ f(x,y,\dots)=y,\dots$
belong to $M$, since the corresponding Zhegalkin
polynomials
$x_i,\ y_i,\dots$ have weight one.
Suppose that some functions
$f(x,y,\dots)$ and $g(x,y,\dots)$ belong to $M$,
i.e.
$$
(f(x,y,\dots))_i=F(x_i,y_i,\dots;x_{i-1},y_{i-1},\dots;\dots),
\quad
(g(x,y,\dots))_i=G(x_i,y_i,\dots;x_{i-1},y_{i-1},\dots;\dots),
$$
where $F$ and $G$ are Zhegalkin polynomials of weight at most one and
without free term.
Then
$$
\eqalign{
&(f(x,y,\dots)\+g(x,y,\dots))_i=
F(x_i,y_i,\dots;x_{i-1},y_{i-1},\dots;\dots)\+
G(x_i,y_i,\dots;x_{i-1},y_{i-1},\dots;\dots)
}
$$
and the weight of this Zhegalkin polynomial
is at most one, i.e.
$f\+g\in M$.
According to $(2'')$,
The $i$th bit of the function $f+g$
is
$$
(f+g)_i=
F\+G\+
F'G'\+
F'F''G''\+F''G'G''\+\dots,
$$
where the polynomial $H'$ is obtained from
$H=H(x_i,y_i,\dots;x_{i-1},y_{i-1},\dots;\dots)$
by the shift of all bits:
$$
H'(x_i,y_i,\dots;x_{i-1},y_{i-1},\dots;\dots)=
H(x_{i-1},y_{i-1},\dots;x_{i-2},y_{i-2},\dots;\dots).
$$
The weight of $H'$
is at most half of the weight of $H$.
Thus, the weight of
$$
F\+G\+F'G'\+F'F''G''\+F''G'G''\+\dots
$$
is at most one and
$f+g\in M$.

\medskip

The remaining part of this section is the proof of the second
assertion.

A \emph{multiple commutators of complexity $n$} is
a formal expression in variables $x,y,\dots$
defined
by induction
as the follows:
\item{}
each variable is a multiple commutator
of complexity 1;

\item{}
an expression $[u,v]$ is a multiple commutator of complexity $n$ if
the expressions $u$ and $v$ are multiple commutators and the sum of their
complexities is $n$.

\enditem
An obvious induction shows that a multiple commutator vanishes if
at least one of the involving variables vanishes.

The \emph{depth} $d(w)$ of a multiple commutator $w$ is also
defined by induction:
\item{}
$d(x) = 0$ if $x$ is a variable;
\item{}
$d([u,v]) = \max(d(u), d(v)) + 1$.

\enditem
For instance, the multiple commutator
$[[x,y],[[z,t],x]]$ has complexity 5 and depth 3.

\Lemma 1.
The $i$th bit of a multiple commutator $w$
vanishes if $i<d(w)$.

\Proof
We use induction on depth. For depth 0, the assertion is true.
Suppose that $d(u)$
lower bits of a multiple commutator $u$
and
$d(v)$
lower bits of a multiple commutator $v$
vanish.
Then formula (2) implies
that $\max(d(u),d(v))+1$ lower bits
of $[u,v]$ vanish, as required.

\Lemma 2.
The depth of a multiple commutator of complexity
at least
$2^n$ is at least
$n$.

\Proof
We use the induction on $n$.
A multiple commutator of complexity 1
(i.e. a variable) has depth 0.
A multiple commutator $w$ of complexity $\ge2^n$, where $n\ge1$,
has the form $w=[u,v]$.
At least one from the multiple commutators $u$ or $v$ has complexity
$\ge 2^{n-1}$ (otherwise, the complexity of $w$ would be less
than $2^n$). By the induction hypothesis, the depth of this multiple
commutator is at least $n-1$. This implies that the depth
of $w$ is at least $n$ by the definition of depth.

\Lemma 3.
In $A_q$, all multiple commutators of complexity $\ge q$ vanish.

\Proof
By Lemma 2, the depth of such a multiple commutator is at least
$\log_2 q$ and, therefore, all bits of this multiple
commutator vanish by Lemma 1.

\medskip
\noindent
{\bf Proof of theorem 1$'$.}
It is sufficient prove that any expression
$$
(2^{k_{1}}x)\odot(2^{k_{2}}x)\odot
\dots\odot
(2^{l_{1}}y)\odot(2^{l_{2}}y)\odot
\dots,
\qqbox{where}
2^{-k_{i,1}}+2^{-k_{i,2}}+\dots+2^{-l_{i,1}}+2^{-l_{i,2}}+\dots\le1,
$$
is
expressible via $\+$ and $+$.
Let us prove a more
strong assertion:
{\sl
any expression of the form
$$
f=(2^{k}u)\odot(2^{l}v)\odot(2^{m}w)\odot\dots,
\qqbox{where
$2^{-k}+2^{-l}+2^{-m}+\dots\le1$
and $u,v,w,\dots$ are multiple commutators}
\eqno(3)
$$
is expressible via $\+$ and $+$ (and variables).
}

Suppose the contrary. Then
there exists an expression of the form (3)
inexpressible via $\+$ and $+$ and such that the inequality
becomes the equality:
$$
2^{-k}+2^{-l}+2^{-m}+\dots=1,
\eqno(4)
$$
Indeed,
$1-(2^{-k}+2^{-l}+2^{-m}+\dots)$ is
a fraction of the form $s\over2^k$, where $k$ is the maximum
of numbers
$k,l,m,\dots$, and, hence, the expression
$$
f\odot\underbrace{(2^{k}u)\odot(2^{k}u)\odot\dots\odot(2^{k}u)}_
{s\;factors}
$$
gives the same function as $f$, but the inequality transforms into
the equality. Note that $2x=[x,x]$ and, hence, $2^ku$ is a multiple
commutator provided $u$ is a multiple commutator.

Let us choose among all nonalgebraic
(not expressible via $\+$ and $+$)
expressions (3) satisfying the equality (4),
all expressions with minimal number
of factors. Among all such minimal-length expressions,
we choose an expression with
the maximal sum of complexities of
commutators $u,v,w,\dots$. Such an expression $f$ exists by Lemma 3.

The number of factors in this expression
is at least two,
because a multiple commutator
can be expressed via $\+$ and $+$ by definition.
Equality (4) implies that the two largest
among exponents $k,l,m,\dots$ are equal.
Let us assume that $k=l$.

Using identity $(2')$,
we obtain
$$
\eqalign{
(2^ku)\odot(2^kv)&=(2\cdot2^{k-1}u)\odot(2\cdot2^{k-1}v)=
\cr&=
[2^{k-1}u,2^{k-1}v]\+(2[2^{k-1}u,2^{k-1}v])\odot
(2\odot2^{k-1}u)\+(2[2^{k-1}u,2^{k-1}v])\odot(2\cdot2^{k-1}v)=
\cr&=
2^{k-1}[u,v]
\+
(2^{k}[u,v])\cdot(2^{k}u)
\+
(2^{k}[u,v])\odot(2^{k}v)
\cr}
$$
Therefore, expression (3) is the
sum of three terms:
$$
f=
\Big((2^{k-1}t)\odot(2^{m}w)\odot\dots\Big)
\+
\Big((2^{k}t)\odot(2^{k}u)\odot(2^{m}w)\odot\dots\Big)
\+
\Big((2^{k}t)\odot(2^{k}v)\odot(2^{m}w)\odot\dots\Big),
\qbox{where $t=[u,v]$}.
$$
All three terms satisfy equality (4).

The first term is algebraic, because its
length
(the number of factors) is less than the length of the initial
expression~$f$, whose
length
is minimal among all nonalgebraic
expressions (3) satisfying equality (4).

The second and third terms
have the same length as $f$, but their total complexity
is higher
(because the complexity of
$t=[u,v]$ is one higher than
the sum of complexities of $u$ and $v$).
Therefore, they are also
algebraic by the choice of~$f$.
Thus, the expression $f$ is algebraic as the sum
of three algebraic terms. This contradiction completes
the proof of Theorem 1$'$ (and of Theorem 1).

\s
3. Proof of Theorems 3 and 2

To prove Theorem 3, note that
the algebra $\Z_q$ with the operations $\+$ and $\o$ is, indeed, a
nilpotent commutative nonassociative ring.
The commutativity of the multiplication $\o$ is obvious;
the distributivity
of the multiplication with respect to the addition $\+$ is obvious too.
The nilpotency also holds:
$((\dots(x\o y)\o z)\o\dots=0$ if the number of
factors is at least $\log_2 q$.
Note that the nilpotency index is generally large than
$\log_2 q$, but it is at most $q$,
i.e. any product of~$q$ elements (with any arrangements of brackets)
vanishes. This can be shown similarly to Lemma 2
(the depth is at least logarithm of the length
for any arrangements of brackets).

The multiplication
$x\o y=2(x\odot y)=(2x)\odot(2y)$
is
expressible via $+$ and $\+$ by Theorem 1$'$.
It remains to prove that the addition $+$
is expressible via the ring operations
$\+$ and $\o$.

Note that $+$ can be expressed via commutator and $\+$
(by the definition of the commutator):
\newline
${x+y=x\+y\+[x,y]}$. Therefore, it is sufficient
to express
commutator via $\+$ and $\o$.

\Lemma 4.
For any positive integer $k$,
the commutator $[x,y]$ can be written in the form
$$
[x,y]=f_k(x,y)\+
\underbrace{
[x,y]\o(x\+y)\o(x\+y)\o\dots\o(x\+y)
}_{k+1\; factors}
,
\eqno{(5)}
$$
where $f_k$ is a polynomial \(in the sense of multiplication $\o$ and
addition $\+$\).
\rm Henceforth, we assume that, in multiple products,
all brackets are
shifted to the left, for example,
$a\o b\o c\o d\:=((a\o b)\o c)\o d$.

\Proof
If $k=1$, then the
required decomposition follows from the identity $(2')$:
$$
[x,y] = x\o y\+[x,y]\o(x\+y).
\eqno{(6)}
$$
For larger $k$, we use induction:
having identity (5) for some $k$, we substitute
identity~(6) in
it the right-hand side of~(5) and obtain
$$
\eqalign{
[x,y]&=f_k(x,y)\+(x\o y\+[x,y]\o(x\+y))\o(x\+y)\o(x\+y)\o\dots\o(x\+y)=
\cr&=
\underbrace{
f_k(x,y)\+x\o y\o (x\+y)\o(x\+y)\o\dots\o(x\+y)
}_{f_{k+1}(x,y)}
\+
\underbrace{
[x,y]\o
(x\+y)\o(x\+y)\o(x\+y)\o\dots\o(x\+y)
}_{k+2\; factors}
},
$$
as required.

\medskip

Applying Lemma 4 for $k=\log_2 q$ and using the nilpotency of the ring,
we obtain an expression of the commutator via $\+$ and $\o$, namely,
$[x,y]=f_{\log_2 q}(x,y)$ that completes the proof of Theorem 3.

\smallskip

Theorem 2 follows immediately from Theorem 3 and the following well-known
fact.

\Theorem
{\rm (see [BaOl88])}.
Each nilpotent ring has a finite basis of identities.

\Remark.
The proof of the existence of a
finite basis for the identities
of nilpotent
rings shows that all identities of such a ring follows
from the identities involving at most $n$ variables, where $n$ is the
nilpotency index, i.e. a number such that all products of $n$ elements
(with any arrangements of brackets) vanish.
This implies the following
fact.

\Corollary 2.
All identities of the algebra $A_q$ follows from the identities involving
at most $q$ elements. There exists an algorithm that, for any given
$q=2^\kappa$,
write out a finite basis of identities of $A_q$.

This basis consists of the addition tables (for $+$ and $\+$) of the free
algebra $F_{q,q}$.

\baselineskip10pt

\REFERENCES

\[GA91]
Artamonov V. A., Salii V. N., Skornyakov L. A., Shevrin L. N.,
Shulgeifer E. G.
General algebra, V.2, Moscow, Nauka. 1991.

\[BaOl75]
Bakhturin Yu. A., Olshanskii A. Yu.
Identical relations in finite Lie rings,
Mat. Sb., 96(138):4 (1975),  543-559.
English translation in
Mathematics of the USSR-Sbornik, 1975, 25:4, 507-523

\[BaOl88]
Bakhturin Yu. A., Olshanskii A. Yu.
Identities.
Algebra-2,
Itogi Nauki i Tekhn. Ser. Sovr. probl. Mat. Fund. Napr, 18,
VINITI, Moscow, 1988, 117-240.
English translation:
Identities,
Algebra, II.
Encyclopaedia of Mathematical Sciences,
Current Problems in Mathematics, Fundamental Directions,
vol. 18, Springer-Verlag, Berlin, 1991,
pp. 117-240.

\[Belo99]
Belov A. Ya.
On non-Specht varieties,
Fund. i Prikl. Mat., 5:1 (1999), 47-66.

\[VaZe89]
Vais A. Ya., Zelmanov E. I.
Kemer's theorem for finitely generated Jordan algebras,
Izv. Vuzov. Ser. Mat., 1989, 6, 63-72.
English translation in
Soviet Math (Izv. VUZ) 33:6 38--47.

\[Grish99]
Grishin A. V. 
Examples of T-spaces and
T-ideals over a field of characteristic two without the finite basis
property,
Fund. i Prikl. Mat., 5:1 (1999),  101-118.

\[Zaits78]
Zaitsev M. V.
On finitely based varieties of Lie algebras,
Mat. Sb., 106(148) (1978), 499-506.
English translation in
Mathematics of the USSR-Sbornik, 1979, 35:2, 165-171

\[Keme87]
Kemer A. R.
Finite basis property of identities of associative algebras,
Algebra i Logika, 26:5 (1987), 597-641.
English translation in
Algebra and Logic,
26:5 (1987), 362-397,

\[Kras90]
Krasilnikov A. N.
The identities of a group with
nilpotent commutator subgroup are finitely based,
Izv. Acad. Nauk SSSR. Ser. Mat., 54:6
(1990), 1181-1195.
English translation in
Mathematics of the USSR-Izvestiya, 1991, 37:3, 539-553

\[Laty73]
Latyshev V. N.
On some varieties of associative algebras,
Izv. Acad, Nauk SSSR. Ser. Mat., 37:5 (1973), 1010-1037.
English translation in
Mathematics of the USSR-Izvestiya, 1973, 7:5, 1011-1038

\[Lvov73]
Lvov I. V.
On varieties of associative rings, I,
Algebra i Logika, 12 (1973), 269-297.
English translation in
Algebra and Logic 12, no. 3 (1973), 150-167

\[Neum69]
Neumann H. Varieties of groups. Springer-Verlag, 1967.

\[Olsh89]
Olshanskii A. Yu.
Geometry of defining relations in groups.
Moscow, Nauka, 1989.
English translation:
Geometry of defining relations in groups.
Mathematics and its Applications (Soviet Series),
vol. 70, Kluwer Academic Publishers Group, Dordrecht, 1991

\[Shch99]
Shchigolev V. V.
Examples of infinitely based T-ideals,
Fund. i Prikl. Mat., 5:1 (1999),  307-312.

\[GuKr03]
Gupta C.K., Krasilnikov A. N.
The finite basis question for varieties of groups -- Some recent results,
Illinois Journal of Mathematics, 47:1-2 (2003), 273.


\[Kras09]
Krasilnikov A. N.
A non-finitely based variety of
groups which is finitely based as a torsion-free variety.
Journal of Group Theory 2009 12:5 , 735-743.

\[Kruse73]
Kruse R. L.
Identities satisfied by a finite ring,
J. Algebra, 26 (1973), 298-318.

\[OaPo64]
Oates S, Powell M. B.
Identical relations in finite groups,
J. Algebra 1 (1964),. 11-39.

\[Speht52]
Specht W.
Gesetze in Ringen. I,
Math. Z., 52 (1950), 557-589.

\end